\documentclass[twocolumn]{aastex6}

\begin{document}

\title{A Hot Companion to a Blue Straggler in NGC188 as Revealed by the Ultra-violet Imaging Telescope (UVIT) on ASTROSAT}

\author{Subramaniam Annapurni$^{1}$, Sindhu N.$^{1,2}$, Tandon S.N.$^{1,3}$, Rao N.Kameswara$^{1}$, Postma J.$^{4}$, 
C\^ot\'e Patrick$^{5}$, Hutchings J.$^{5}$, Ghosh S.K.$^{6,7}$, George K.$^{1}$,
Girish V.$^{8}$, Mohan R.$^{1}$, Murthy J.$^{1}$, Sankarasubramanian K.$^{1,8,9}$, Stalin C.S.$^{1}$, Sutaria F.$^{1}$, Mondal C.$^{1}$ \and Sahu S.$^{1}$}
\affil{$^1$Indian Institute of Astrophysics, Koramangala II Block, Bangalore-560034, India\\
$^2$School of Advanced Sciences, VIT University, Vellore 632014, India\\
$^3$Inter-University Center for Astronomy and Astrophysics, Pune, India\\
$^4$University of Calgary, Canada\\
$^5$National Research Council, Canada\\
$^6$National Centre for Radio Astrophysics, Pune, India\\
$^7$Tata Institute of Fundamental Research, Mumbai, India\\
$^8$ISRO Satellite Centre, HAL Airport Road, Bangalore 560017\\
$^9$Center of Excellence in Space Sciences India, Indian Institute of Science Education and Research (IISER), Kolkata,
Mohanpur 741246, West Bengal, India}
\email{purni@iiap.res.in}

\begin{abstract}
We present early results from the Ultra-Violet Imaging Telescope (UVIT) onboard the ASTROSAT observatory.  We
report the discovery of a hot companion associated with one of the blue straggler stars (BSSs) in the old
open cluster, NGC188.  Using fluxes measured in four filters in UVIT's Far-UV (FUV) channel, and two filters 
in the near-UV (NUV) channel, we have constructed the spectral energy distribution (SED) of the star WOCS-5885, after 
combining with flux measurements from GALEX, UIT, UVOT, SPITZER, WISE and several ground-based facilities. The resulting
SED spans a wavelength range of 0.15~${\mu}$m  to 7.8~${\mu}$m. This object is found to be one of the brightest FUV 
sources in the cluster.  
An analysis of the SED reveals the presence of two components.
The cooler component is found to have a temperature of 6\,000$\pm$150~K, confirming that it is a BSS. Assuming it to be a
main-sequence star, we estimate its mass to be $\sim$ 1.1 - 1.2M$_\odot$. The hotter component, with an estimated temperature
of 17\,000$\pm$500~K, has a radius of $\sim$ 0.6R$_\odot$ and L $\sim$ 30L$_\odot$.
Bigger and more luminous than a white dwarf, yet cooler than a sub-dwarf, we speculate that it is a post-AGB/HB star that
has recently transferred its mass to the BSS, which is known to be a rapid rotator.
This binary system, which is the first BSS with a post-AGB/HB companion identified in an open cluster, is an ideal laboratory to
study the process of BSS formation via mass transfer. 

\end{abstract}


\section{Introduction} \label{sec:intro}
Old open clusters are important tools for understanding the advanced stages of stellar evolution for both
single and binary stars.  Such clusters are well studied in optical and near-IR pass bands, 
but only a handful have been studied in the Ultraviolet (UV). In the old open clusters M67, NGC188 and 
NGC6791, populations of hot stars were studied by Landsman et al. (1998) using images from the Ultraviolet 
Imaging Telescope (UIT).  Analysis of wide-field UV photometry of nearby clusters M67, NGC188, NGC2539 
and M79 obtained with the Swift Ultraviolet Optical Telescope (UVOT) showed that UV 
bright stars can be readily identified using the UV color magnitude diagram (CMD; Siegel et~al. 2014). 

Blue straggler stars (BSSs) are cluster members that are brighter,
and bluer, than stars on the upper main sequence (Sandage 1953). 
The BSS population of the open cluster NGC188, in addition to a few sub-giants and yellow giants, comprises
approximately 25\% of the evolved cluster population (see Mathieu \& Geller 2015, and references therein). 
BSSs are believed to have gained mass by some process and hence have a rejuvenated lifetime on the main-sequence. 
The BSS are  found in a wide variety of environments, such as
open clusters, globular clusters, Galactic field and in elliptical galaxies.
Suggested formation mechanisms include mergers in hierarchical
triples (Perets \& Fabrycky 2009), collisions during
dynamical encounters (Knigge et al. 2009; Leigh \& Sills 2011),
and mass transfer (Chen \& Han 2008) on the red giant or asymptotic
giant branches (RGB and AGB; respectively). Recently,
Gosnell et al. (2014, 2015) used HST to study BSSs in NGC188 in the UV region, detecting
white dwarf (WD) companions to seven BSS and thereby confirming that at least some BSSs
can form by mass transfer. Three of their BSSs were found to have cool WD companions; four
 were found to have hotter WDs. 
To date, though, no BSS belonging to an open cluster has
been found that possesses a post-AGB/RGB star as a companion.


NGC188 is a well studied open cluster thanks to its richness, high metallicity and relatively old age. The 
age of the cluster is determined to be 7 Gyr, with a reddening of E(B$-$V) = 0.09$\pm$0.02 mag 
and a distance of 2047 pc (Sarajedini et al. 1999), though  some variations are found (See Hills et al. (2015)).
Proper motion studies from the WIYN Open Cluster Study (WOCS; Platais et al. 2003) suggest that 1050
stars are members of the cluster.  A  total of 24 BSS candidates were 
cataloged by Ahumada \& Lapasset (2007), with Geller et al.
(2008) showing 20 BSSs to be confirmed members based on their radial velocities. In their 
UIT study, Landsman et al. (1998) identified three stars with exceptionally blue colors.

In this paper, we report the detection of a hot companion to the star WOCS-5885 using data from the
Ultra-Violet Imaging Telescope (UVIT). Below, we introduce the
instrument, describe the data and analysis methods, followed by a discussion of results.

\section{UVIT data}
The UVIT instrument contains two 38-cm telescopes: one for the far-ultraviolet (FUV) region (130--180~nm); 
and the other for the near-ultraviolet (NUV; 200--300~nm) and visible (VIS) regions (320--550~nm) ranges; these 
are divided using a dichroic mirror for beam-splitting. UVIT is primarily an imaging instrument, simultaneously 
generating images in the FUV, NUV and VIS channels over a 28$^\prime$-diameter circular 
field.{\footnote{A low-resolution, slit-less spectroscopic mode is also available in FUV and NUV channels.}}
Each channel can be divided into smaller pass bands using a selectable 
set of filters. Full details on the
telescope and instruments can be found in Tandon et al. (2016) and Subramaniam et al. (2016), along with
the initial calibration results.  The primary photometric calibration for all FUV filters and two of the 
NUV filters were performed using observations of HZ4, a white dwarf spectrophotometric standard star. Note
that the present calibrations for the UVIT filters cover the central region of $\sim$ 7arcmin. 

NGC188 was observed as UVIT's ``first light" object on 30 November 2015. The cluster
was observed every month to track the variation in  UVIT sensitivity over the first six months of 
its operation. In this paper, we present the calibrated flux of WOCS-5885, located near the center of the UVIT 
field, in four FUV and two NUV pass bands. All data used in the analysis were obtained on 
26 January 2016. 
The star studied here is repeatedly observed and we have detections within the central 5 arcmin. 
The variations in the count rates suggest the sensitivity variation to be about 2-3\%, calibration error of 2-3\% and  
the photon noise to be 1-2\%, with a cumulative effect of about 5\%. In the case of the specific filter, B15, we noticed that
the above estimates are almost double, which sets the error to be 10\%.

Images were corrected for distortion, flat field illumination and spacecraft drift using the customised 
software package CCDLAB (Postma et al. 2017, in preparation). Aperture photometry was performed 
on the images to estimate the counts after correcting for the background and saturation effects.
The filters, their effective wavelength, the unit conversion factor, and estimated fluxes and errors
are recorded in Table~1.  

\setcounter{table}{0}
\begin{table}[h!]%
\centering
\caption{UVIT flux measurements for WOCS-5885. Columns 1-5 record the filters; the effective wavelength in \AA;
the unit conversion factor for each filter; the measured flux (in units of $ergs~cm^{-2}~s^{-1}~\AA^{-1}$); and its associated error. 
The measured flux is the product of the count rate and the unit conversion for each filter (see Tandon et al. 2016 for details).} \label{tab:flux}
\begin{tabular}{|c|c|c|c|c|}
\hline
Filter & $\lambda_{eff}$ & Unit conv & Flux &Error \\
\hline
FUV channel & & & & \\
\hline
F148W (CaF2-1) & 1480.8 & 0.292E-14 & 5.09E-14 & 0.25E-14 \\
F154W (BaF2)   & 1540.8 & 0.345E-14 & 4.73E-14 & 0.19E-14 \\
F169M (Sapphire) & 1607.7& 0.428E-14 & 4.09E-14 & 0.20E-14 \\
F172M (Silica) & 1716.5 & 0.106E-13 & 3.51E-14 & 0.18E-14\\
\hline
NUV channel & & & & \\
\hline
N219M (B15) & 2195.5 & 0.545E-14&2.06E-14&0.21E-14\\
N279N (N2)  & 2792.3 & 0.364E-14 & 1.19E-14 &0.06E-14\\
\hline
\end{tabular}
\end{table}

\begin{figure}[]
\figurenum{1}
\plotone{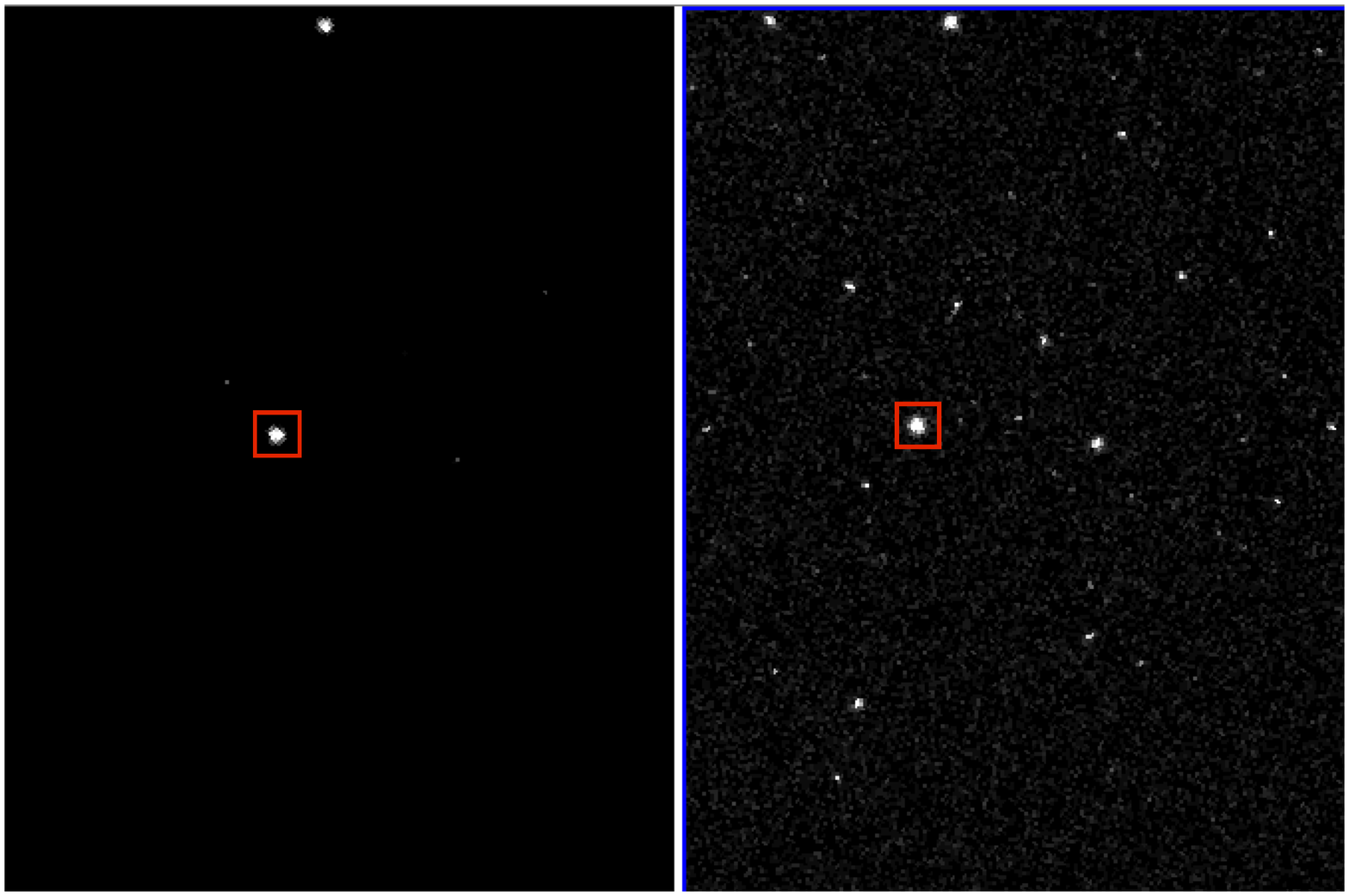}
\caption{FUV (left) and NUV (right) images of NGC188 taken in the CaF2 and N2 filters respectively. WOCS-5885 is marked by the red square.
The images obtained on 18 February 2016 are aligned with North up the East towards the left.}
\end{figure}

\begin{figure}[]
\figurenum{2}
\plotone{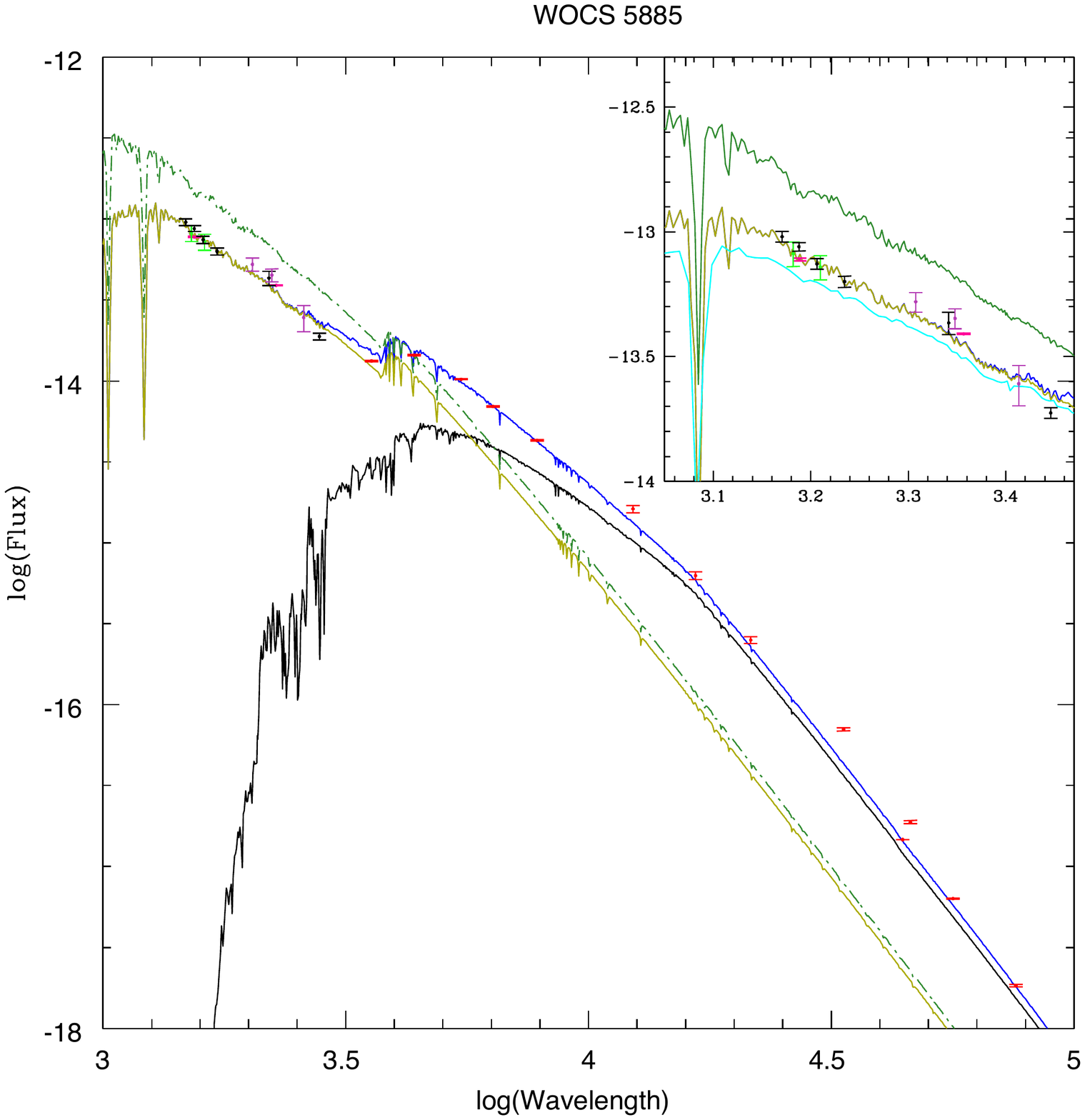}
\caption{The extinction-corrected spectral energy distribution (SED) of WOCS-5885. The black (UVIT), magenta (GALEX) and
green (UIT), pink (UVOT) points indicate the UV fluxes (shown in the inset as well); all other flux measurements are shown in red.  Kurucz Model
spectra (Log(g) = 5.0) for the separate components are shown in gold (17\,000~K) and black (6\,000~K), with the composite spectrum in blue. For comparison,
we have shown a hotter spectrum of temperature (20\,000~K) in dark green. We have also shown the helium rich
model spectrum for a temperature of 16\,000~K, $\log{g} = 4.0$, H= 0.30, He =0.70, and CN = 0.00005 from
Jeffery et al. (2001), in Cyan.
Scaling factors of 4.45E-22  and 3.1E-23 have been used to combine the 6\,000~K 
and 17\,000~K spectra, respectively.  The unit of wavelength is \AA~ and flux is $ergs~cm^{-2}~s^{-1}~\AA^{-1}$.
The data points which were not considered for the final reduced $\chi^2$ value are, GALEX(NUV), U, W1, W2, I2, I3.}
\end{figure}
\begin{figure}[]
\figurenum{3}
\plotone{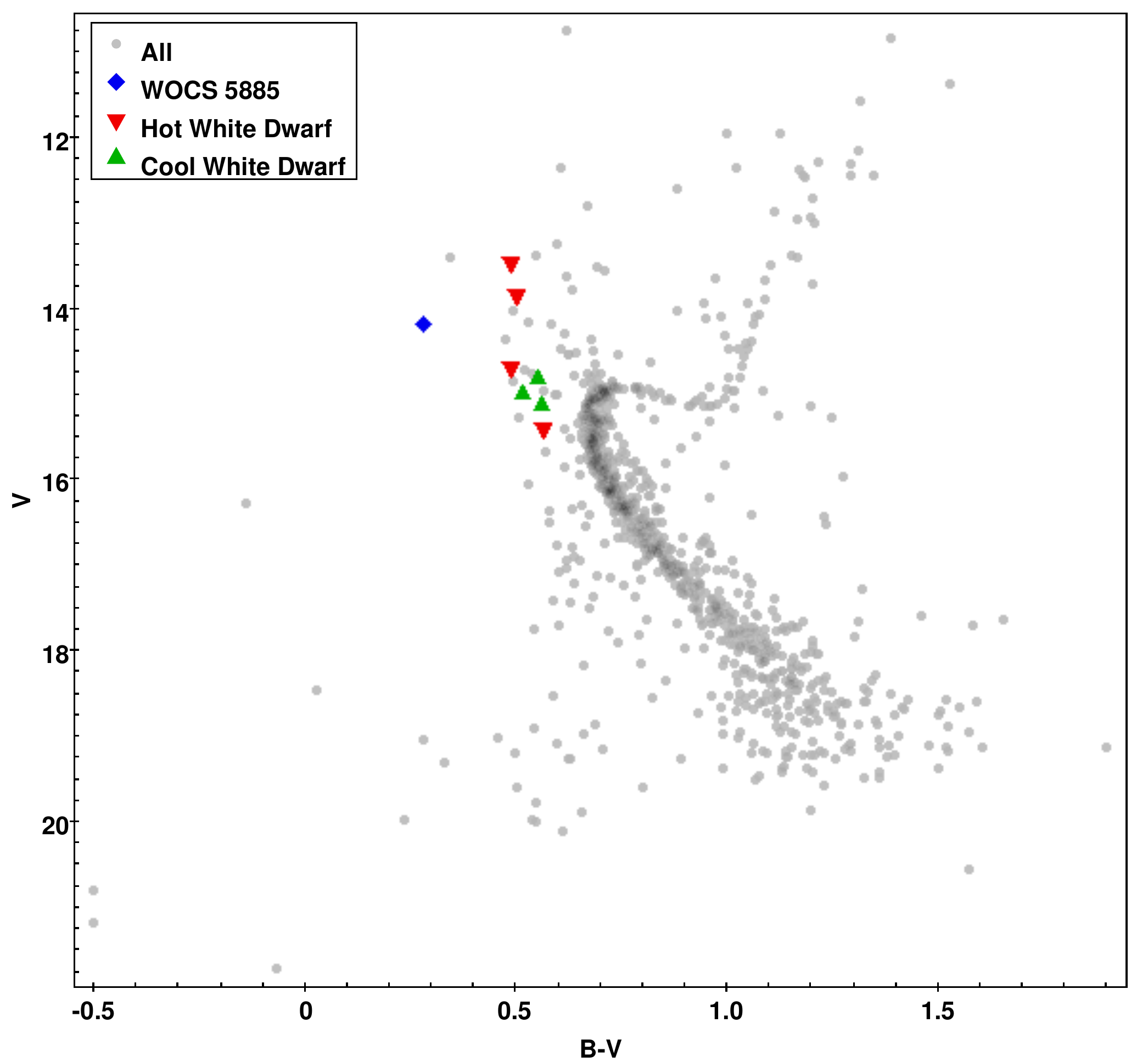}
\caption{Color-magnitude diagram (CMD) of NGC188 showing the location of WOCS-5885 and the seven BSS+WDs 
sources studied by Gosnell et al. (2014). The members (proper motion and radial velocity) of the cluster are shown in the
CMD, using the optical photometry of Sarajedini et al. (1999).}
\end{figure}

\section{Spectral Energy Distribution}

We describe the spectral energy distribution (SED) for the BSS, WOCS-5885 (Table 2), 
located in the inner region of NGC188. This is known to be a single-lined spectroscopic binary of 
uncertain membership (Geller et al. 2008). 
They found the star to be a rapid rotator and mentioned that they could not
establish membership and designated it as an uncertain member due to a lack of
measurements.  Stetson et al. (2004) found a membership probability of 53\% (SMV-8422) and Dinescu et al. (1996) 
assigned a probability of 80\% (D-702, found to be an extended object). 
It is considered to be a BSS based on its location in the CMD. Landsman et al.(1998) mentioned that this
may have a composite spectrum, as suggested by its colors, with a possibility of a very hot star and a BSS as components.
Siegel et al. (2014) suspected this to be a red giant and a pre-WD binary. 

\setcounter{table}{1}
\begin{table}[h!]%
\centering
\caption{The basic parameters of WOCS-5885. The coordinates and binarity are taken from Geller et al. (2008);
magnitudes, reddening and distance from Sarajedini et al. (1999);
membership probabilities from Stetson et at. (2004) and Dinescu et al. (1996).  }
\begin{tabular}{|c|c|}
\hline
RA (2000) (hh:mm:ss) & 00:48:20.19\\
Dec (2000)(dd:mm:ss) & +85:13:27.1\\
V (mag) & 14.13 \\
(B$-$V) (mag) & 0.25 \\
Membership prob & 53\%, 80\% \\
Binary & Yes \\
Reddening E(B$-$V) & 0.09 mag\\
Distance (pc) & 2047 \\
\hline
Estimated parameters &\\
\hline
{\bf Cool component} & \\
Temperature & 6000($\pm$150)~K\\
Mass  & $\sim$ 1.1 - 1.2M$_\odot$\\
Radius  & $\sim$ 1.1 - 1.6R$_\odot$\\
Luminosity  & $\sim$ 1.4 - 2.8 L$_\odot$\\
{\bf Hot component} &\\
Temperature & 17000 ($\pm$ 500)~K\\
Radius & $\sim$0.6 R$_\odot$\\
Luminosity  & $\sim$ 30 L$_\odot$\\
\hline
\end{tabular}
\end{table}

Our UVIT images clearly show the star to be brighter in the FUV than in the NUV, and much fainter in optical 
images. The star is not found to have any significant elongation or any extended feature in the UVIT images (figure 1). In order to characterise the temperature and the evolutionary status of the star, we created a 
multi-wavelength SED using UVIT data, supplemented by literature data. We combined flux 
measurements from GALEX (FUV and NUV), UIT (2 pass bands), SWIFT-UVOT (3 pass bands, Siegel et al. 2014), 
UVBRI (Sarajedini et al. 1999), JHK (2MASS),
I2,I3,I4 (Spitzer) and W1,W2 (WISE). {\footnote {We did not include the W3 and W4 data due to larger PSF and low S/N. We obtained the UVOT magnitudes from Siegel et al. (2014), from their figure 5. The magnitude in uvw1 as found from the bottom two figures differ by about 0.2 magnitudes, we adopted the mean value. The adopted magnitudes and errors are uvw2=15.0$\pm$0.1, uvm2=15.0$\pm$0.1, uvw1=15.1$\pm$0.2.}}
The final SED consists of 26 data points spanning the
wavelength range of 0.15--7.80~$\mu$m (Figure 2).
It can be seen from Figure~2 that the UVIT fluxes are in good agreement 
with those estimated by UIT, UVOT and GALEX. 

We used the virtual observatory tool, VOSA (VO SED analyzer; Bayo et al. 2008) which allows the user to perform 
a statistical test to compare the observed data with the selected model. The tool uses the filter transmission curve 
to calculate the synthetic photometry of the theoretical model which is compared with the observed flux. The tool 
also corrects the observed flux for extinction in the respective band (Fitzpatrick 1999, Indebetow et al. 2005) and 
then SED fitting with the model for a selected star is performed. We adopted a reddening of E(B$-$V)= 0.09. To
extinction correct the UVIT and UIT fluxes, we used the reddening law of Cardelli, Clayton, \& Mathis (1989).
The reduced $\chi^2$ value is estimated 
using the expression $$\chi^2 = \frac{1}{N-4}\sum_{i=1}^{N} \Big\{ \frac{(F_{o,i}-F_{m,i})^2}{\sigma_{o,i}^2} \Big\}$$
where $F_{o,i}$ is the observed flux, $F_{m,i}$ is the scaled model flux and $\sigma_{o,i}^2$ is the error in the observed flux.
The synthetic flux values of the UVIT filters were computed by us, whereas for other filters, we used the values
computed by VOSA, except for the UIT filters. 
The rise in flux for UV wavelengths can be clearly seen in Figure~2, suggesting the
presence of a hot component. In order to fit the SED, we need spectra which have coverage over the
UV to IR wavelength range. We used Kurucz models (Castelli et al. 1997) because the spectra need both a large
wavelength and temperature coverage. Although Next-Gen models could also be used, the study by Subramanian 
et al. (2016) suggested that the Kurucz models
are preferred for temperatures hotter than 3\,500~K. 

We tried initially to fit the full SED with a single temperature but found this approach to yield unsatisfactory results.
The UV part of the SED seems to suggest a relatively hot temperature, whereas the IR part suggests a cooler temperature.
Therefore, we created a composite spectrum comprising of two components. 
We fitted the UV range and the optical-IR range separately using the VOSED. The temperatures and errors are
estimated by the tool. The combination that best fits the
observed SED (with the least $\chi^2$ value) was found to have temperatures of 17\,000($\pm$500)~K and 6\,000($\pm$150)~K. 
The reduced  $\chi^2$ value using 24 data points are 87 (12), 2698 (1169)
and 7582 (868) for composite, cool and hot components respectively.  
The reduced $\chi^2$ values for well-fitting 18 data points, identified by the tool, followed by a visual inspection are shown in the paranthesis.
In Figure 2, we also show the SED fitted with two spectra,
along with a composite spectrum. For comparison, we have also shown a model spectrum of temperature 20\,000K. The N279N flux and the U band flux are lower,
whereas the WISE fluxes are higher with respect to the fitted spectrum.
The N279N filter is centered on the Mg II spectral line and line absorption could be a possible reason for the reduced flux.
Although the temperatures are reliable,  the derived $\log{g}$ ($\sim$ 5.0) are not very accurate since the range of values available in the models is limited.
In order to compare with special models, such as helium rich models for hot stars, we have shown the model LTE spectrum for a 
temperature of 16\,000~K, from Jeffery et al. (2001), in figure 2. The Helium-rich model shows a reduction in flux in U band, which may support the observed reduction in the U band flux. In general, both the models appear comparable.

The high temperature of the hotter component (17\,000$\pm$500~K) is clearly revealed by the UVIT data, especially in 
the flux measured using the F148W and F154W filters; inset in Figure 1 suggests that the UV flux is rising at least until 
1481~\AA. ~The data also confirm that the temperature of the hot component is not high enough to be classified as a 
sub-dwarf, as demonstrated by the slope of the FUV part of the SED.
The estimated temperature of the cooler component confirms that it is indeed a BSS, as it is similar to the temperature of other BSSs in 
NGC188 (Table~1; Gosnell et al. 2015) and higher than the turn-off temperature of NGC 188 ($\sim$ 5500~K, based on the turn-off color). 
The optical region of the SED is fitted well by the combined flux from the two components.
The Padova models (Marigo et al. 2008) suggest a mass of 1.1 - 1.2 M$_\odot$ for a star of similar temperature on the MS. 
This is also similar to the mass of other BSS in NGC 188 (Perets 2015). 

In obtaining the composite spectrum, 
the flux of each component is scaled by a  factor which is equal to $(R/D)^2$ where $R$ is the radius and $D$ is the 
distance. We estimated the radii of the component stars, using the relation $R/R_\odot = (T_\odot/T)^2(L/L_\odot)^{(0.5)}$,
the luminosity is estimated from the absolute magnitude, bolometric correction and a distance of 2kpc. 
For the BSS, we also estimated the Luminosity using the relation given in table 3 of Eker et al. (2015) and Padova models.
The BSS is found to have a radius of 
$\sim$ 1.1 - 1.6R$_\odot$, and the Luminosity of $\sim$ 1.4 - 2.8L$_\odot$, whereas the hot component is found to have a radius of $\sim$ 0.6R$_\odot$,
and a Luminosity of $\sim$ 30L$_\odot$.
Hence, the hot component is much larger than a WD (R $\sim$ 0.01-0.02R$_\odot$, Tremblay et al. 2016 and references therein). The hot star is about 3 times more luminous than hot B sub-dwarfs (T$_{eff}$ $>$ 27000~K  L $\sim$ 10 L$_\odot$) and blue horizontal branch stars (L $\sim$ 10 L $_\odot$)    
(Heber 2016). Also, the FUV luminosity is found to be much higher than
those of the BSS+WD of similar temperature (e.g., WOCS-4540, WOCS-5379) as estimated by Gosnell et al. (2015).
These two observations suggest that the hot component cannot be a WD or a sub-dwarf, therefore we speculate that it is a 
post-AGB star or a post-HB star (depending on the initial mass of the star). Thus, WOCS-5885 is likely to be an example of
the rare class of BSS+post-AGB/HB binary star. This also suggests that the BSS acquired mass quite recently from the post-AGB/HB star - a scenario that is supported by its rapid rotation,  as the BSS can get spun up due to mass transfer (Ivanova 2015). Mathieu \& Geller (2009) found
that many NGC 188 BSSs are rotating more rapidly than the main-sequence stars and speculated that the rapid rotation of blue stragglers may place upper limits on their ages. 

As the BSS has a mass of 1.1 -1.2M$_\odot$ and the turn-off mass of NGC 188 is $\sim$ 1.0M$_\odot$, the mass gained by the BSS is at least 0.1-0.2M$_\odot$. The mass of the progenitor of the post-AGB/HB is likely to be slightly more massive to evolve first. Recently, Milliman, Mathieu \& Schuler (2015) suggested that mass transfer
from AGB stars to be the dominant formation mechanism for BSSs in NGC 188. The
evolution of the primary star, once the mass transfer starts, is complicated and is addressed by a few authors for a few
specific cases (reviews by Sills (2015), Ivanova (2015)). As there are no similar objects observed or modelled, we are
unable to compare or comment on the estimated parameters of the hot component. 
In this study, we have considered the two components to be physically associated.
In the direction of NGC 188, approximately one hot star is expected per sq. degree, as per figure 10 of Bianchi et al. (2011).
As there are at least 3 hot stars in this cluster field suggesting an over density, the hot component is likely to be associated with the cluster.

\section{Discussion}
We have presented  early science results from UVIT that demonstrate its UV imaging capabilities. 
Observations in four filters in the FUV and two filters in the NUV channel were used.
The UVIT flux clearly demonstrates the rising flux of the hot component in the FUV bands; these measurements 
allow accurate  temperature estimates from SED fitting. The estimated fluxes are based on the 
available calibrations and will be improved as calibrations continue. The UVIT fluxes are in good agreement with those
from UIT, GALEX and UVOT. We therefore demonstrate that accurate flux measurements can be carried out with UVIT
in the 0.150-0.30~$\mu$m wavelength range. 

We have combined our UVIT data with flux measurements from other missions, as well as from ground based optical/IR telescopes, to demonstrate that WOCS-5885 could be a rare BSS+post-AGB/HB binary --- the first of its kind to be 
identified in an open cluster. WOCS-5885 is probably the progenitor of systems composed of a BSS and hot-WD 
companion, like those identified by Gosnell et al. (2015). The temperature of the companion rules out the possibility 
this has a red-giant along with a pre-WD, as suggested by Siegel et al. (2014). Figure~3 shows the location of this star,
along with the other BSS+WDs in this cluster, in the cluster color-magnitude diagram. WOCS-5885 is the relatively 
blue object while the five BSS+WD binaries are redder in comparison. 

Recently, Milliman, Mathieu \& Schuler (2015) identified a few
BSSs in NGC 6819 to be enhanced in Barium, likely to be  formed through mass 
transfer from an AGB star, but none of these stars are found to be binaries. Sivarani et al. (2004) studied a possible field BSS and
found large overabundances of  s-process elements, due to 
accreted material from a companion, formerly an AGB star. Ryan et al. (2002) studied rapidly rotating lithium 
deficient stars that are possible field BSSs. WOCS5885 is unique such that this is a BSS where a possible post-ABG star is
found as a companion, a target for testing the above findings.
This is an interesting target 
for  BSS surface composition studies and possible chemical signatures of recent accretion events. 
The physical parameters of the components can thus be used to constrain mass transfer models of BSS formation (Chen \& Han 2008). 

\section{Conclusions}

Our analysis of UVIT imaging for the open cluster NGC188 has allowed us to reach the following conclusions 
concerning WOCS-5885, a likely BSS and confirmed UV-bright source belonging to the cluster:
\begin{enumerate}
\item We find the star to be a certain binary and use UVIT to characterize its two components.
\item The cooler component (6\,000$\pm$150~K) is found to be a BSS, whereas the hot component (17\,000$\pm$500~K) is
speculated to be a post-AGB/HB star based on its luminosity and radius. We therefore argue that WOCS-5885 is likely to be a BSS+post-AGB/HB binary.
\item The mass of the BSS is estimated to be  $\sim$ 1.1-1.2M$_\odot$. The initial mass of the donor star is estimated to be $\ge$ 1.2M$_\odot$.
\item This rare system is the first of its kind to be 
identified in an open cluster is an ideal candidate to study the chemical composition of the BSS and constrain the theories of BSS formation via mass transfer.
\end{enumerate}

\acknowledgments
We thank the referee for a careful reading of the manuscript and very thoughtful comments which improved the paper. Sindhu acknowledges support from IIA for short term visit.
UVIT project is a result of collaboration between IIA, Bengaluru, IUCAA, Pune, TIFR, Mumbai, several centres of ISRO, and CSA. Indian Institutions and the Canadian Space Agency have contributed to the work presented in this paper. Several groups from ISAC (ISRO), Bengaluru, and IISU (ISRO), Trivandrum have contributed to the design, fabrication, and testing of the payload. The Mission Group (ISAC) and ISTRAC (ISAC) continue to provide support in making observations with, and reception and initial processing of the data.  We gratefully thank all the individuals involved in the various teams for providing their support to the project from the early stages of the design to launch and observations with it in the orbit.

\end{document}